\documentclass{article}

\usepackage[letterpaper,top=2cm,bottom=2cm,left=3cm,right=3cm,marginparwidth=1.75cm]{geometry}

\usepackage{amsmath}
\usepackage{graphicx}
\usepackage{graphicx}
\usepackage{dcolumn}
\usepackage{bm}

\usepackage{graphicx} 
\usepackage{array}
\usepackage{multirow}
\usepackage{booktabs}
\usepackage{rotating}
\usepackage{tabu}
\usepackage{relsize}
\usepackage{amssymb}
\usepackage{pifont}
\usepackage{bbding}
\usepackage{starfont}
\usepackage{wasysym}
\usepackage{eqnarray}
\usepackage{tablefootnote}
\usepackage{ragged2e}
\usepackage[numbers]{natbib}
\usepackage{xcolor}
\usepackage{amsmath}

\definecolor{indigo1}{rgb}{0 , 0.4470,  0.7410}
\definecolor{orange1}{rgb}{0.8500 0.3250 0.0980}
\definecolor{yellow1}{rgb}{0.9290, 0.6940, 0.125}
\definecolor{purple1}{rgb}{0.4940, 0.1840, 0.5560}
\definecolor{green1}{rgb}{0.4660, 0.6740, 0.1880}
\definecolor{blue1}{rgb}{0.3010, 0.7450, 0.9330}
\definecolor{crimson1}{rgb}{0.6350, 0.0780, 0.1840}
\definecolor{lavender1}{rgb}{0.996, 0.812, 0.996}
\definecolor{brown1}{rgb}{0.6050, 0.4023, 0.2344}
\definecolor{teal1}{rgb}{0, 0.847, 0.847}
\definecolor{darkblue1}{rgb}{0, 0.0664, 0.6055}
\definecolor{peach1}{rgb}{1, 0.69, 0.486}
\definecolor{gray1}{rgb}{0.8, 0.804, 0.776}
\definecolor{darkgreen1}{rgb}{0.349, 0.506, 0.322}
\definecolor{red1}{rgb}{0.89, 0.141, 0.169}

\definecolor{redplot}{rgb}{0.043    0.067    0.443}
\definecolor{purpleplot}{rgb}{0.447    0.169    0.416}
\definecolor{blueplot}{rgb}{0.827    0.129    0.176}

\title{Dynamics of fluid-driven fractures in the viscous-dominated regime}
\author{Sri Savya Tanikella$^1$, Marie C. Sigallon$^{1,2}$, Emilie Dressaire$^{1}$\\
1. University of California Santa Barbara\\
2. Ecole Polytechnique, CNRS, Institut Polytechnique de Paris}

\begin{document}
\maketitle

\begin{abstract}
During hydraulic fracturing, the injection of a pressurized fluid in a brittle elastic medium leads to the formation and growth of fluid-filled fractures. A disc-like or penny-shaped fracture grows radially from a point source during the injection of a viscous fluid at a constant flow rate. We report an experimental study on the dynamics of fractures propagating in the viscous regime. We measure the fracture aperture and radius over time for varying mechanical properties of the medium and fluid and different injection parameters. Our experiments show that the fracture continues to expand in an impermeable brittle matrix, even after the injection stops. {In the viscous regime, the fracture radius scales as $t^{4/9}$ during the injection. Post shut-in, the crack continues to propagate at a slower rate, which agrees well with the predictions of the scaling arguments, as the radius scales as $t^{1/9}$. The fracture finally reaches an equilibrium set by the toughness of the material.} The results provide insights into the propagation of hydraulic fractures in rocks.
\end{abstract}

\section{Introduction}
Hydraulic fracturing is a well-stimulation technique used to recover natural gas and oil from reservoirs with low permeability, such as shale formations. The US Environmental Protection Agency reports 
that the natural gas production from hydraulically fractured wells in the United States saw a 10-fold increase between 2000 and 2015 \cite{1}. 
The formation of fractures in rocks has other applications, including carbon sequestration and geothermal energy extraction \cite{1b, 1c}. 

It is estimated that almost 1 million wells have been hydraulically fractured since the 1940s. 
As hydraulic fracturing has become more prevalent, so has the need to characterize the associated risk to the local environment and populations. Over the past two decades, groundwater contamination and induced seismicity have been linked to hydraulic fracturing operations. Between 2000 and 2013, there was at least one hydraulically fractured well system within 1 mile of the water sources of 3900 public water systems in the continental United States \cite{2}. The water from these systems was distributed to more than 8.6 million people year-round in 2013. Another major concern is the induced earthquakes associated with hydraulic fracturing \cite{3}.
%\cite{Yu2021}
 Earthquakes of magnitude 4.0 can be caused by the disposal of wastewater in fractured reservoirs \cite{4,5,6,7}.
 The associated risks increase as the distance between stimulated wells and groundwater wells or fault lines decreases. Therefore, understanding the dynamics of a fracture during and after the fluid injection is critical for risk assessment.

When a pressurized Newtonian fluid is injected from a point source into a uniform impermeable brittle matrix, a disk-like hydraulic fracture forms and propagates. This penny-shaped fracture results from the coupling of three mechanisms: (1) the elastic deformation of the fracture surfaces, (2) the propagation of the fracture at the rim of the fluid-filled region, and (3) the flow of the fluid in the fracture. Theoretical and numerical modeling of the penny-shaped crack has been developed since the seminal work of Sneddon $\&$ Mott \cite{8}. 
Yet the fracture dynamics remain complicated to model owing to the multi-scale nature of the problem, \cite{9,10}, with viscous dissipation associated with fluid transport through the volume and the stress concentration at the tip of the fracture. In limiting regimes, in which the viscous dissipation or the fracture opening controls the dynamics of the fracture, tip asymptotes can be defined \cite{11,12,13,14,15,16}. 

In the viscous-dominated or zero-toughness regime, the elastic stresses drive the radial fracture propagation, which is limited by the viscous stresses. In the toughness-dominated regime, the stress concentration or stress intensity factor controls the fracture expansion, and the viscous stresses are negligible. 
To study the two-propagation regimes, laboratory-scale experiments use hydrogels, whose brittle elastic properties are analogous to those of rocks. For example, gelatin is a clear material that allows fracture visualization for a wide range of mechanical properties \cite{17,18}. 
  Upon injection of an aqueous solution or oil from a needle, a penny-shaped fracture forms at the injection point and expands radially through the gelatin, perpendicularly to the needle. Radius and aperture measurements in experimental model systems agree with the theoretical predictions during the fluid injection. \cite{18,19,20,21}.%

The dynamics of a fracture after the injection, i.e., post-shut-in, differs depending on the propagation regime {\cite{221}}. Here we assume that the fracture propagates in a mobile equilibrium with no fluid lag. The influence of gravity on fracture propagation is negligible \cite{22}. In the toughness-dominated regime, the fluid-filled fracture propagates when the stress intensity factor is equal to the toughness of the matrix. When the fluid injection stops, the elastic pressure is no longer sufficient to sustain the propagation of the fracture. In the toughness regime, the final geometry of the fracture is reached when the injection stops. In the viscous regime, however, the material's toughness does not limit the propagation of the fracture. The viscous dissipation in the fluid balances the elastic stresses during and after the injection. This study focuses on the dynamics of fractures that propagate in the viscous-dominated regime during and post-injection. {Previous experimental studies have investigated the fracture dynamics during the injection of a viscous fluid in a hydrogel matrix \cite{18,21} and the closure of the fracture post-injection in porous materials like cement and plaster with a typical Young's modulus of the order of 1 GPa \cite{231}. Those studies characterized the decline in pressure inside the fracture post-shut-in, where the pressure decrease is due to various phenomena such as leak-off. Numerical studies predicted fracture growth and pressure reduction after shut-in in non-permeable matrices \cite{232,233}}. Here, we conduct injection experiments in gelatin blocks to study the post shut-in dynamics of a fracture in the viscous regime. The fracture initially forms and propagates during the injection of a viscous Newtonian liquid. The injection is stopped, and we record the time dependence of the fracture radius and aperture until both properties become constant, indicating that the fracture has reached its equilibrium configuration.  We observe that the fracture created in the viscous regime continues to propagate even after the injection stops. Consequently, we identify three different regimes of propagation in our experiments: (1) propagation during injection, (2) propagation post shut-in or at constant volume, and (3) saturation. {To the best of our knowledge, the experiments presented here are the first observations for the three regimes in a hydrogel matrix, and the data agree well with the corresponding scaling laws \cite{221,232,233}.}  

This paper is structured as follows. In \S2, we discuss the experimental set-up and methods and our observations. In \S3, we summarize the scaling arguments and derivation of the dimensionless parameters. The experimental results and theoretical predictions are compared in \S4. Our conclusions are summarized in \S5. 

\section{Experiments and observations}
To study the fracture dynamics in the viscous-dominated regime, we inject a viscous liquid in a high Young's modulus and low toughness gelatin block. Using dyed silicone oil, we can characterize the geometry of the fracture, i.e., its radius and aperture, as a function of the radial distance from the injection point during and after the injection. 

\subsection{Experimental methods}
\begin{figure}
\includegraphics[width = 0.9\linewidth]{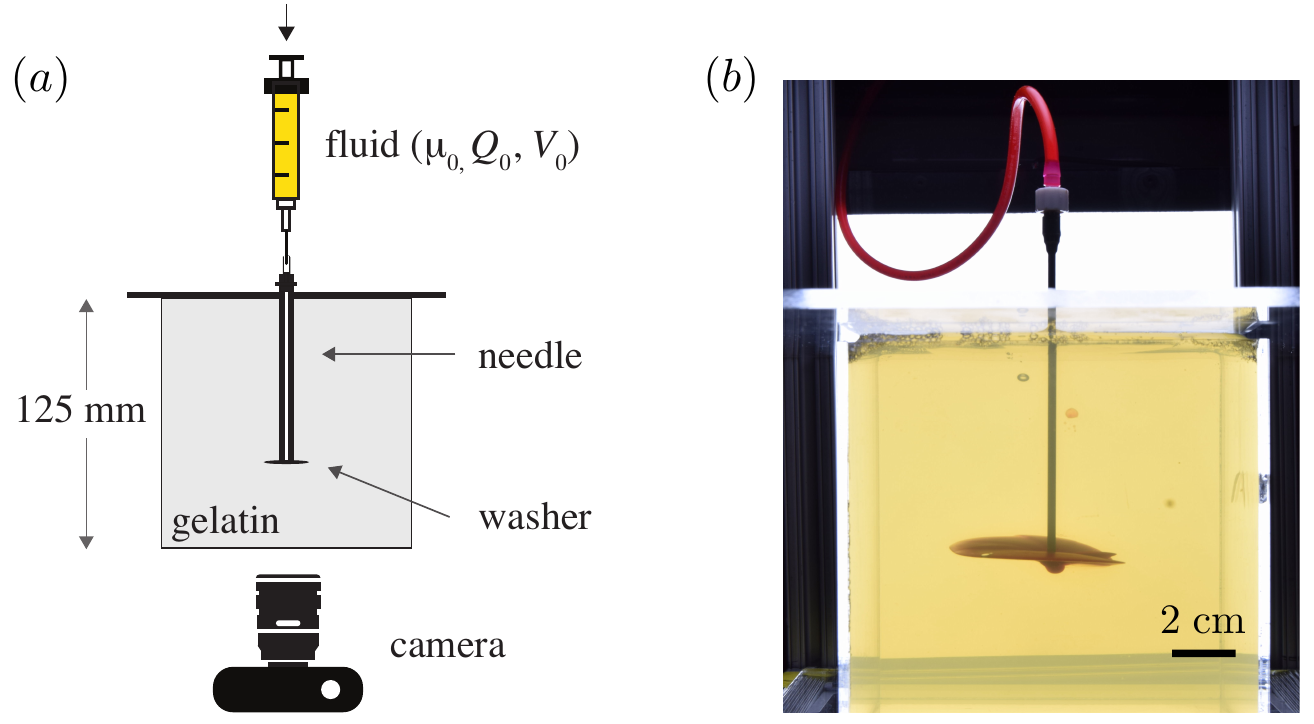}
\caption{Experimental set-up: (a) schematic and (b) penny-shaped fracture formed by injecting silicone oil of viscosity $\mu = 10$ Pa.s at a flow rate $Q_0 = 10$ ml.min$^{-1}$) into a block of gelatin with Young's modulus $E = 88$ kPa. The oil is dyed with red oil-soluble food color.}
\label{fig:setup}
\end{figure}
The gelatin is prepared by heating ultra-pure water to $60^{o}$C and slowly adding gelatin powder (Gelatin type A; Sigma-Aldrich, USA) while mixing. The gelatin is then allowed to be set over 24 hours at room temperature in a cubic clear container (12.5 cm $\times$ 12.5 cm $\times$ 12.5 cm) around a blunt needle as represented in figure \ref{fig:setup}. The Young's modulus of the gelatin $E$ is measured with cylindrical samples of height and diameter equal to $2.5$ cm. The cylinders are tested under compression using a custom-built displacement-controlled load frame. The Young's moduli range between $88$ to $144$ KPa ${\pm 10 \%}$ for mass fractions of gelatin powder in water between $20$ - $25 \%$. The fracture energy and Poisson's ratio of the gelatin are assumed constant with $\gamma_S \approx 1$ J.m$^{-2}$ and $\nu \approx 0.5$ respectively \cite{23}. The inner diameter of the blunt needle is equal to $2.15$ mm. A plastic washer of diameter of about 6 mm is placed at the tip of the needle to ensure the propagation of the fracture in a plane perpendicular to the needle and parallel to the lens of the camera. We inject silicone oils of different viscosity $\mu$ to fracture the gelatin. Viscosity measurements are conducted using an MCR 92 Anton Parr rheometer with a parallel plate measuring system. The values obtained at $20^o$C are listed in table \ref{tab:exp} and have an error of $\pm 1\,\%$. We use a syringe pump (KDS Legato 200 series infusion syringe pump) to inject the fracturing fluid at a controlled flow rate $Q_0$ ranging from $5$ to $28$ ml.min$^{-1}$. The injection stops when a volume $V_0$ of fluid has been injected. \\
{To ensure that the experiments are in the viscous regime, we estimate the ratio of the toughness-related pressure and the viscous pressure. This ratio is called the dimensionless toughness ($\mathcal{K}_s$), and its maximum value is reached at the time of shut-in \cite{221}}

\begin{equation}
\mathcal{K}_s = K^{\prime} \frac{t_s^{1 / 9}}{E^{\prime 13 / 18} \mu^{\prime 5 / 18} Q_o^{1 / 6}},
\end{equation}
{with the effective viscosity $\mu' = 12\mu$, the effective toughness
$K' = 4\left(\frac{2}{\pi}\right)^{\frac{1}{2}}K_{IC}$, the toughness $K_{IC}= \sqrt{2 \gamma_S E'}$ and the effective Young's modulus $E' = E/\left(1-\nu^2\right)$. The dimensionless toughness is of order 1 for all experiments as presented in table 1. We, therefore, expect the fracture propagation to be limited by the viscous dissipation associated with the fluid flow in the fracture}.\\
The silicone oil is dyed using oil-based food color to help visualize the propagation of the fracture. The list of experiments and the corresponding parameters are summarized in table \ref{tab:exp}.
The propagation of the fracture is recorded using a Nikon D5300 camera with a Phlox\textsuperscript{\tiny\textregistered} LED panel ensuring uniform backlighting. The images are processed using a custom-made MATLAB code to determine the radius of the fracture $R$.

\begin{table}[]
\centering
\begin{tabular}{@{}lllllllllllll@{}}
\toprule
Exp. &  &   Markers         &  & E (kPa)   &  &  $\mu$ (Pa.s) &  & $Q_{0}$ (ml.min$^{-1}$)  &  & $V_{0}$ (ml) & 
 &  $\mathcal{K}_s$  \\ \midrule
1    &  & $\color{yellow1} {\mathlarger{\mathlarger{\bullet}}} \hspace{-2.3 mm} \color{black} {\mathlarger{\mathlarger{\mathlarger{\circ}}}}$ / $\color{black} {\mathlarger{\mathlarger{\mathlarger{\circ}}}}$ &  & 88  &  & 10.3 &  & 15 &  & 6  &  &  1.6 \\
2   &  & \textcolor{orange1}{\rotatebox[origin=c]{90}{\large$\blacktriangle$}} \hspace{-4.9mm} \raisebox{-1.5pt}{\rotatebox{90}{$\bigtriangleup$}} / \raisebox{-1.5pt}{\rotatebox{90}{$\bigtriangleup$}} &  & 88  &  & 10.3 &  & 5  &  & 6  &  &  2.17  \\
3    &  & ${\color{black} \hspace{0 mm} {\mathlarger{\mathlarger{\blacktriangle}}} }$ ${\hspace{-4 mm} \raisebox{-0.5pt}{{$\bigtriangleup$}}}$ \hspace{-1.2 mm} /   ${\hspace{-0.4 mm} \color{black} {\raisebox{-0.5pt}{{$\bigtriangleup$}}}}$ &  & 88  &  & 10.3 &  & 10 &  & 6   &  &  1.79  \\
4    &  & \textcolor{purple1}{\rotatebox[origin=c]{-90}{\large$\blacktriangle$}} \hspace{-5.5mm} \raisebox{7.5pt}{\rotatebox{-90}{$\bigtriangleup$}} / \raisebox{7.5pt}{\rotatebox{-90}{$\bigtriangleup$}}  &  & 88  &  & 10.3 &  & 25 &  & 4.3  &  &  1.34 \\
5    &  & \textcolor{green1}{\rotatebox[origin=c]{180}{\large$\blacktriangle$}} \hspace{-5.5mm} \raisebox{7.5pt}{\rotatebox{180}{$\bigtriangleup$}} / \raisebox{7.5pt}{\rotatebox{180}{$\bigtriangleup$}} &  & 88  &  & 10.3 &  & 28 &  & 4.3   &  &  1.30\\
6    &  & $\color{blue1} {\mathlarger{\blacklozenge}}  \hspace{-3.4 mm} \color{black}$ {\raisebox{-0.0pt}{{$\lozenge$}}} / {\raisebox{-0.0pt}{{$\lozenge$}}}  &  & 88  &  & 10.3 &  & 15 &  & 8   &  &  1.65 \\
7    &  & ${\color{crimson1} {\hspace{-0.5 mm}{\mathlarger{\mathlarger{\mathlarger{\times}}}}}}\hspace{-3.5 mm}\color{black} \square$  \hspace{-1.5 mm} / ${\color{crimson1} {\hspace{-0.6 mm}{\mathlarger{\mathlarger{\mathlarger{\times}}}}}}$ &  & 88  &  & 10.3 &  & 15 &  & 4   &  &  1.53 \\
8    &  & $\color{black} \blacksquare \hspace{-2.65mm}\color{black} \square$ / $\square$ &  & 88  &  & 20 &  & 25 &  & 3   &  &  1.06 \\
9    &  & ${\color{brown1} {\hspace{-0.2 mm}{\mathlarger{\mathlarger{\mathlarger{+}}}}}}\hspace{-3.4mm}\color{black} \square$  \hspace{-1.4 mm} / ${\color{brown1} {\hspace{-0.6 mm}{\mathlarger{\mathlarger{\mathlarger{+}}}}}}$ &  & 116 &  & 10.3 &  & 10 &  & 4    &  &  1.61\\
10    &  & $\color{blue1} \blacksquare \hspace{-2.65mm}\color{blue1} \square$ / $\color{blue1} \square$  &  & 144  &  & 10.3 &  & 20 &  & 3   &  &  1.23 \\
11    &  & $\color{red1} \blacksquare \hspace{-2.65mm}\color{red1} \square$ / $\color{red1} \square$  &  & 144  &  & 30 &  & 15 &  & 3.5   &  &  0.996
\end{tabular}
\caption{\label{tab:exp} List of Experiments. The markers on the left and right of the $/$ symbol correspond to the data recorded during and after the injection respectively.}
\end{table}

\subsection{Thickness measurements}
We use the light absorption technique pioneered by Bunger \cite{24} to measure the fracture aperture using a soluble dye in the injected fluid. In our system, white light illuminates the sample and a filter is placed on the camera to measure the light intensity at a single wavelength. The filtered wavelength corresponds to the maximum absorbance of the dye. At this wavelength, the absorbance $A_{\lambda}$ follows Beer's law:
\begin{equation}
    A_{\lambda} = - \log_{10}\left(\frac{I_{\lambda}}{I_{\lambda,0}}\right)= \epsilon_{\lambda} c \, h
\end{equation}
where $I_{\lambda,0}$ is the background intensity and $I_{\lambda}$ is the intensity of light after it passes through a fluid layer of thickness $h$, with a dye concentration $c$.  The fitting parameter $\epsilon_{\lambda}$ depends on the dye-fluid combination and the concentration and is obtained through calibration. Here, the fracturing fluid is dyed with Nile Red at a concentration of 0.2 g.L$^{-1}$, and the wavelength of the optical filter used was 632 nm. The value of the fitting parameter $\epsilon_{\lambda}c= 3.54 \times 10^{-3}$ mm$^{-1}$ is obtained through the calibration process using liquid layers whose thickness ranges from $0.14$ mm to $2$ mm.

\subsection{Observations}
\begin{figure}
\includegraphics[width = \linewidth]{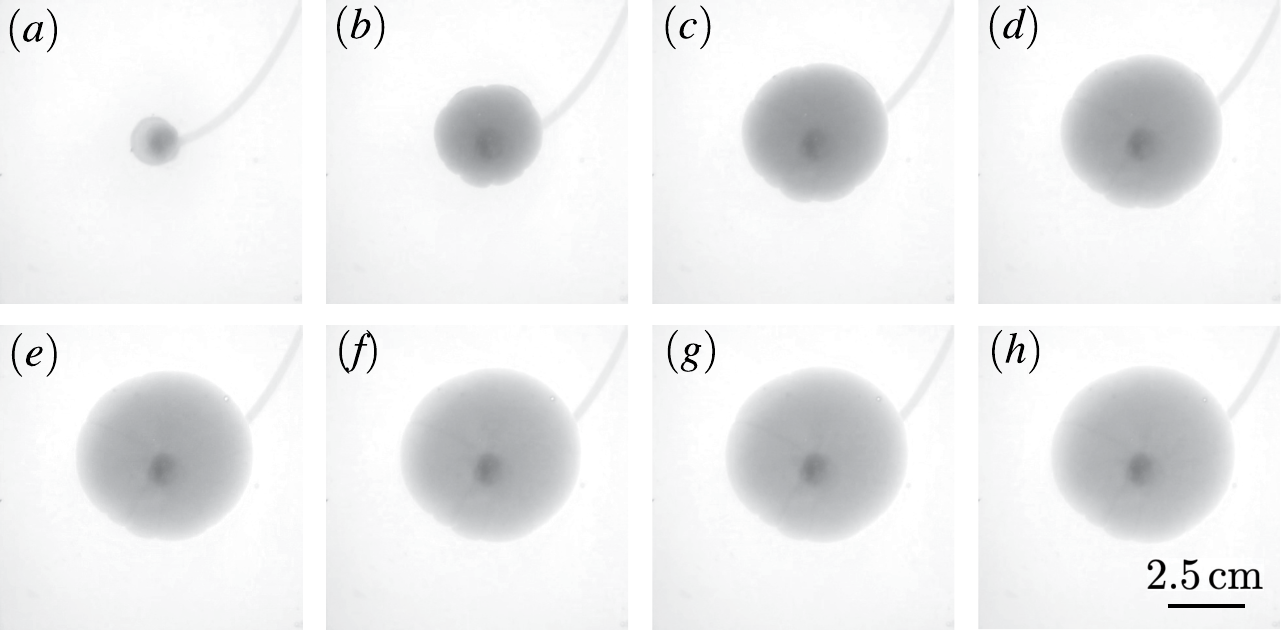}
\caption{Fracture growth during and after the injection. The fluid is dyed with red food color to enhance the contrast. The grayscale images presented here are obtained by filtering the red channel from the color images. The first images are recorded during the injection: (a) $t = 0$ s, (b) $t = 18$ s, (c) $t = 36$ s. The fracture continues to grow after the injection stops (d) $t = 54$ s, (c) $t = 126$ s, (f) $t = 198$ s, (g) $t = 270$ s. The recording ends when the fracture has reached its equilibrium configuration (h) $t = 396$ s. The experimental parameters are: Young's modulus \textit{E} = 88 KPa, flow rate $Q_0$ = 10 ml/min, volume injected $V_0$ = 6 ml and fluid viscosity $\mu$ = 10 Pa.s.}
\label{fig:time_series}
\end{figure}

During the injection process, the fracture forms at the tip of the needle and propagates along the washer and beyond, expanding radially. Figure \ref{fig:time_series} presents a time series of the fracture propagation (see also electronic supplementary material, Movie S1). As the fracture grows radially, its thickness increases as its color becomes darker. When the injection stops, the fracture expands radially at a slower pace. As the fracture grows, its color fades, indicating that the aperture decreases with time. Since the amount of fluid in the fracture needs to be conserved, as the radius increases, the width of the fracture decreases. Finally, the fracture stops growing. To ensure that the finite volume of material and the bounding container walls are not affecting fracture growth, we experimented with a larger volume of gelatin (see Appendix A for more details). The maximum fracture radius is independent of the size of the gelatin block, indicating that the viscous-dominated fracture stops expanding when it reaches equilibrium. 

\section{Scaling arguments}
 
\begin{figure}
\begin{center}
\includegraphics[width = 0.7\linewidth]{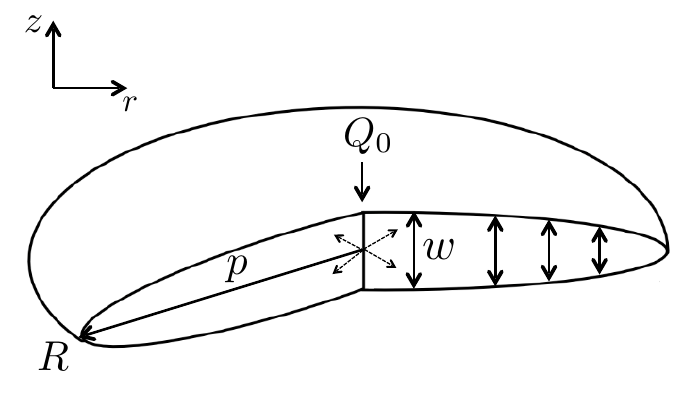}
\caption{Diagram showing the radial fracture geometry.}
\label{fig_sim}
\end{center}
\end{figure}

{The fracture results from the injection of a high-viscosity Newtonian fluid in a brittle elastic matrix that is impermeable. We assume there is no lag between the fracture tip and the fluid front.} The fracture is initially driven by an incompressible fluid of viscosity of $\mu$ pumped at a constant flow rate $Q_0$. The elastic medium is characterized by Young’s modulus $E$, Poisson’s ratio $\nu$, and toughness $K_{IC}$. The injection through a point source leads to the formation and radial propagation of a penny-shaped fracture with no lag between the fluid and the fracture tip, as represented in figure \ref{fig_sim}. The experiments conducted in this study are in the viscous-dominated regime during the injection. The fracture grows as the elastic stresses in the fracture boundaries drive the fluid outward. The viscous dissipation associated with fluid transport in the fracture limits growth. The material toughness is negligible and does not contribute to the fracture dynamics during the injection.

{The evolution of the radius and width of the fracture in the viscous dominated regime can be modeled using scaling arguments, derived for both the injection and post "shut-in" stages of the propagation \cite{13,221}}
We make the following assumptions: (i) the stresses in the matrix are well described by Linear Elastic Fracture Mechanics (LEFM), (ii) the lubrication theory can be used to model the flow; (iii) the fracture propagates continuously in a mobile equilibrium; and (iv) the matrix is impermeable (no leak-off). To describe the fracture aperture $w(r,t)$, radius $R(t)$, and pressure $p(r,t)$, we solve the coupled equations that describe (a) the viscous flow of the fracturing fluid in the time-dependent fracture, see equation (\ref{eq:2.1}), (b) the elastic deformation of the solid material, see equation (\ref{eq:2.2}), (c) the fracture propagation criteria based on LEFM, see equation (\ref{eq:2.3}), and the global mass balance or fluid mass conservation, see equations (\ref{eq:2.4}) or (\ref{eq:2.5}). The net pressure couples these equations. The non-dimensional forms of the equations are summarised below, and a detailed derivation is provided in Appendix B. The equations are non-dimensionalized with $R = R_{o}\hat{R}$, $w = w_{o}\hat{w}$, $P = P_{o}\hat{P}$, and $t = t_{o}\hat{t}$, where $R_{o}$, $w_{o}$, $P_{o}$ and $t_{o}$ represent the characteristic radius, aperture, pressure of the fracture and timescale of the propagation, respectively. {As proposed by Savitski and Detournay} \cite{13}, we define the effective viscosity $\mu' = 12\mu$, the effective toughness
$K' = 4\left(\frac{2}{\pi}\right)^{\frac{1}{2}}K_{IC}$ and the effective Young's modulus $E' = E/\left(1-\nu^2\right)$.
The dimensionless lubrication equation writes \cite{21}:
\begin{equation}
    \frac{\partial \hat{w}}{\partial \hat{t}}=\frac{t_{0} w_{0}^{2} p_{0}}{\mu^{\prime} R_{0}^{2}} \frac{1}{\hat{r}} \frac{\partial}{\partial \hat{r}}\left(\hat{r} \hat{w}^{3} \frac{\partial \hat{p}}{\partial \hat{r}}\right)\label{eq:2.1}.
\end{equation}
The elastic deformation leads to the following:
\begin{equation}
    \hat{w}=\frac{8}{\pi} \frac{p_{0} R_{0} \hat{R}}{w_{0} E^{\prime}} \int_{\hat{r} / \hat{R}}^{1} \frac{\xi}{\sqrt{\xi^{2}-(\hat{r} / \hat{R})^{2}}} \int_{0}^{1} \frac{x \hat{p}}{\sqrt{1-x^{2}}} \mathrm{d} x \mathrm{d} \xi \label{eq:2.2}.
\end{equation}
The fracture propagation requires:
\begin{equation}
    \frac{K^{\prime}}{p_{0} R_{0}^{1 / 2}}=\frac{2^{7 / 2}}{\pi \sqrt{\hat{R}}} \int_{0}^{\hat{R}} \frac{\hat{p}}{\sqrt{\hat{R}^{2}-\hat{r}^{2}}} \hat{r} \mathrm{~d} \hat{r}\label{eq:2.3}.
\end{equation}
Since the gelatin is considered impermeable, the volume of the fracture is equal to the volume of fluid injected. During the injection, the volume of the fracture is equal to
\begin{equation}
    \hat{Q} \hat{t}=2 \pi \frac{R_{0}^{2} w_{0}}{Q_{0} t_{0}} \int_{0}^{\hat{R}} \hat{r} \hat{w} \mathrm{~d} \hat{r}\label{eq:2.4}.
\end{equation}
The injection stops when the fracture volume is equal to $V_0 = Q_{0}t_{0}$. After the injection, the volume of the fracture writes:
\begin{equation}
     \hat{V}=2 \pi \frac{R_{0}^{2} w_{0}}{V_{0}} \int_{0}^{\hat{R}} \hat{r} \hat{w} \mathrm{~d} \hat{r}\label{eq:2.5}.
\end{equation}
To determine the fracture dynamics during and after the injection, we now solve the sub-set of equations relevant to each stage of fracture propagation.

\subsection{Viscous regime: constant flow rate propagation}
To obtain the scaling relations that describe the fracture dynamics during the injection at a constant flow rate, we assume that the major form of dissipation of energy in the fracture is the viscous dissipation of the fluid flow. We set the dimensionless parameters in equations (\ref{eq:2.1}), (\ref{eq:2.2}) and (\ref{eq:2.4}) equal to 1. We recover the scaling relations for the radius $R_{o}$ and aperture $w_{o}$ of the fracture  { originally derived by Savitski and Detournay} \cite{13}:
\begin{eqnarray}
    R_{0} &\approx& \left(\frac{Q_{0}^{3}E't_{0}^{4}}{\mu'}\right)^{1/9} \label{eq:InjR}\\
    w_{0} &\approx& \left(\frac{\mu'^{2}Q_{0}^3t_0}{E'^{2}}\right)^{1/9} \label{eq:InjW}.
\end{eqnarray}
During the injection, both the radius and the aperture are increasing functions of time, which is consistent with our observations.

\subsection{Viscous regime: propagation at constant volume}
Once the injection is complete, the volume inside of the fracture is constant {\cite{221}}. The viscous dissipation of the fluid is assumed to be the limiting factor in the propagation of the fracture. Setting the dimensionless parameters of equations (\ref{eq:2.1}), (\ref{eq:2.2}) to 1 and (\ref{eq:2.5}) and solving for the characteristic radius $R_{0}$ and aperture $w_{0}$, we get:
\begin{eqnarray}
    R_{0} &\approx& \left(\frac{V_{0}^{3}E't_{0}}{\mu'}\right)^{1/9} \label{eq:VR}\\
    w_{0} &\approx& \left(\frac{\mu'^{2}V_{0}^3}{E'^{2}t_{0}^{2}}\right)^{1/9} \label{eq:VW}.
\end{eqnarray}
The radius is still increasing with time, at a slower rate than during the injection. The aperture of the fracture is now a decreasing function of time. This is consistent with the fact that the total volume of fluid in the fracture ($\propto R_0^2w_0$) needs to be conserved, independent of time. 
%%%%%%%%%%%%%%%%%%%%%%%%%%%%%%%%%%%%%%%
\subsection{Toughness regime: saturation}
Based on the scaling relations derived for the viscous regime, the fracture radius is an increasing function of time at constant fracture volume. Yet, the elastic pressure in the fracture is a decreasing function of time, as the wall deformation decreases. Eventually, the material toughness is no longer negligible. Indeed, if the pressure in the fracture is too low, the material no longer fractures and the propagation of the fracture stops. 
We assume that saturation, unlike the previous two stages of propagation, is controlled by the fracture opening or material toughness. The fracture opening criterion as a function of the pressure and stress intensity factor is given by equation (\ref{eq:2.3}). This equation coupled with the elastic stress in the gelatin matrix, equation (\ref{eq:2.2}) and the volume conservation of the fracture, equation (\ref{eq:2.5}) leads to the following scaling {\cite{221}}:
\begin{eqnarray}
    R_{0} &\approx& \left(\frac{V_{0}E'}{K'}\right)^{2/5} \label{eq:ER}\\
    w_{0} &\approx& \left(\frac{K'^{4}V_{0}}{E'^{4}}\right)^{1/5} \label{eq:EW}.
\end{eqnarray}
The radius and the aperture of the fracture are now functions of the volume of fluid and mechanical properties of the matrix.  They are independent of time. 

\section{Results and observations}
In the following section, we report quantitative experimental results, and their comparison with the scaling arguments derived above for both the radius and width of the fracture. 

\begin{figure}[h!]
    \centering
    \includegraphics[width = 1\linewidth]{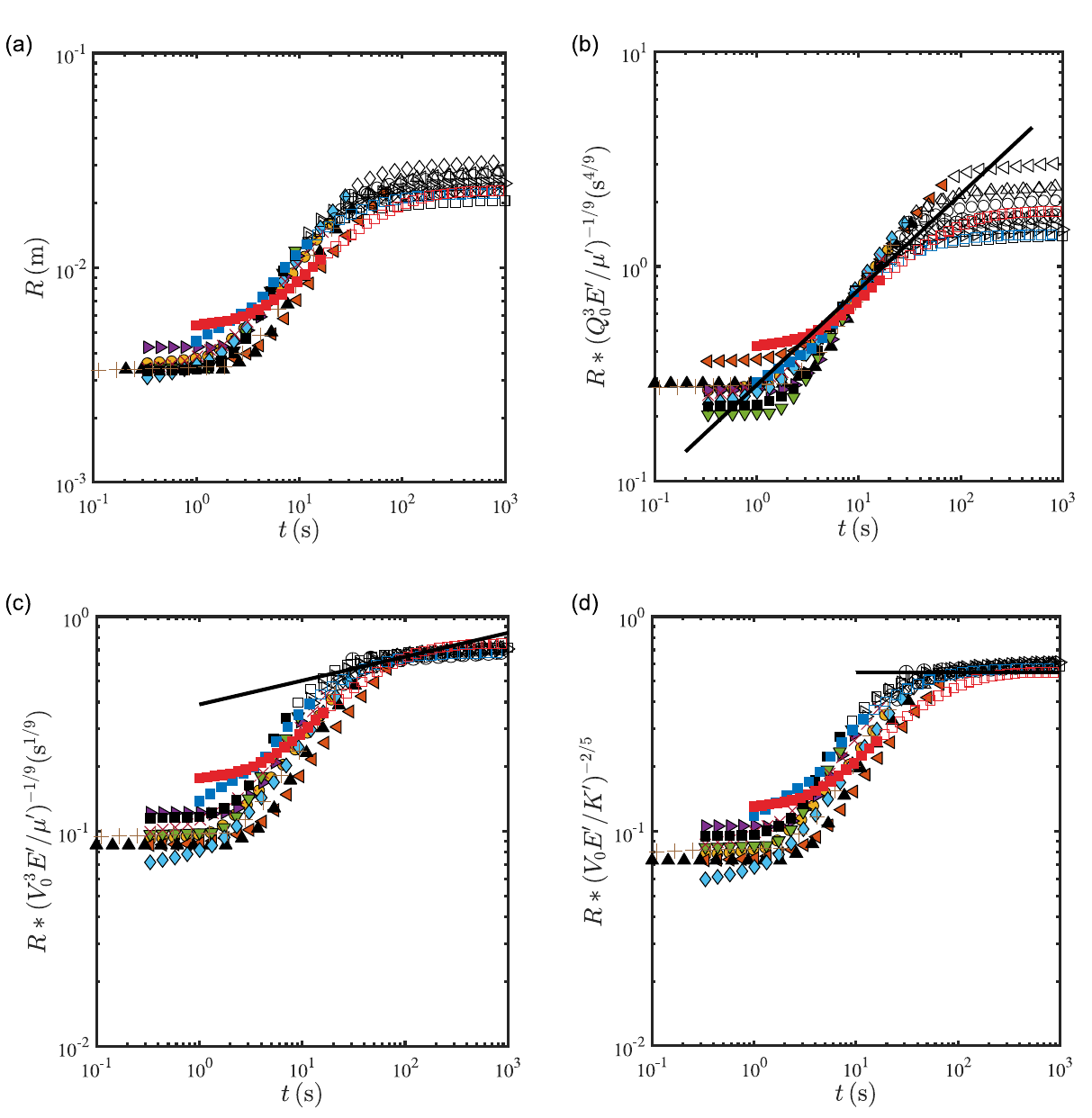}
    \caption{Radius of the fracture formed during experiments 1-9, see table \ref{tab:exp} for corresponding parameters. Data collected during and after the injection are displayed with solid and open symbols, respectively. (a) The radius of the fracture over time. (b) Rescaled radius using equation (\ref{eq:InjR}) for the regime (i) as a function of time. (c) Rescaled radius using equation (\ref{eq:VR}) for the regime (ii) as a function of time. (d)  Rescaled radius using equation (\ref{eq:ER}) for the regime (iii) as a function of time. For each regime, the best fit line is represented by a solid black line.}
    \label{fig:radius}
\end{figure}

\subsection{Radius measurements} 
We first measure the radius of the fracture as a function of time for the experimental parameters summarized in table \ref{tab:exp}. The raw experimental data are shown in figure \ref{fig:radius}(a). The solid symbols correspond to the values recorded during the injection. In contrast, the open markers indicate that the data were recorded after the injection. For some of the experiments, such as experiment $8$, we see a $50\,\%$ increase in the radius of the fracture after the injection stops. 

We successively rescale the data using the scaling laws derived above and obtain figures \ref{fig:radius}(b-d), where the fit line is shown in black. {On the log-log plot, the slope of the line is set to the value of the power-law derived using scaling arguments. The y-intercept is obtained by minimizing the mean squared error and corresponds to the prefactor, which can be predicted theoretically \cite{15}}. The propagation dynamics exhibit three stages described by (i) the viscous-dominated propagation during the injection, (ii) the viscous-dominated propagation at constant volume after the injection stops, and (iii) a toughness-controlled saturation.
For the fracture dynamics during the injection [regime (i) and figure \ref{fig:radius}(b)], we rescale the data using the viscous scaling. After the early times that correspond to the fracture formation and its propagation over the washer, the data points collapse onto a best-fit line which has a prefactor, $k_1 = 0.28$ and an exponent, $\alpha_1 =4/9$ where
\begin{equation}
   R(t)=k_1\left(\frac{E^{\prime} Q_{0}^{3}}{\mu^{\prime}}\right)^{1 / 9} t^{\alpha_1}. \label{eq:4.1} 
\end{equation}
{The theoretical prefactor derived by  Savitski and Detournay \cite{15} is equal to 0.7.} The low value of the prefactor obtained in our experiments is consistent with measurements previously reported for the viscous-dominated regime \cite{18,21}.

In the next regime [regime (ii) and figure \ref{fig:radius}(c)], the fracture fluid continues to propagate after the injection stops, and the open symbols collapse on the best fit line
\begin{equation}
   R(t)=k_2\left(\frac{E^{\prime} V_{0}^{3}}{\mu^{\prime}}\right)^{1 / 9} t^{\alpha_2},\label{eq:4.2} 
\end{equation}
which has a prefactor $k_2= 0.39$ and an exponent, $\alpha_2 = 1/9$. The results indicate that the propagation continues to be dominated by viscous dissipation. For time values above 100 s, the increase in radius no longer follows the best fit line: the radius reaches its maximum or equilibrium value. 

In the saturation regime, regime (iii), and figure \ref{fig:radius}(d)], we rescale the data using the volume of the fracture and the mechanical properties of the gelatin matrix. All data collapse on an average value represented with a horizontal line whose y-intercept $k_3 \approx 0.58$ where,
\begin{equation}
   R(t)=k_3\left(\frac{E^{\prime} V_{0}}{K^{\prime}}\right)^{2 / 5}.\label{eq:4.3} 
\end{equation}
{The experimental prefactor of 0.58 is comparable to the expected theoretical prefactor of 0.85 \cite{221}}. This result indicates that the propagation controlled by the viscous dissipation stops when the stress intensity factor at the tip of the fracture is too low to sustain the formation of the fracture. 

\begin{figure}
    \centering
    \includegraphics[width = 0.75\linewidth]{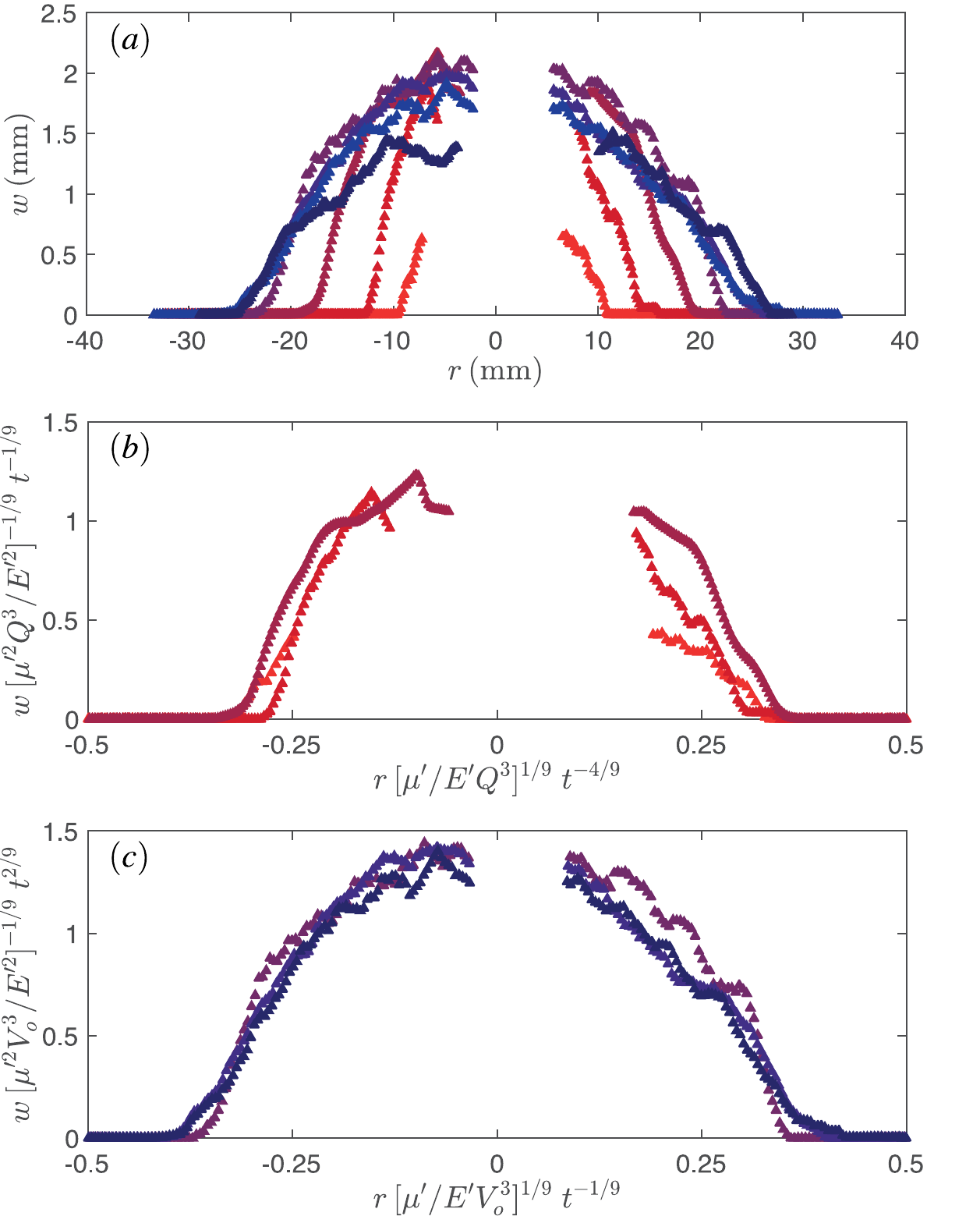}
    \caption{(a) Fracture profiles measured during and after the injection for $t = [9, 21, 33, 81, 105, 117, 762]$ s with time increasing from clear red to dark blue. (b) Rescaled fracture  profiles using the scaling laws derived for the injection in \S 2.1. (c) Rescaled fracture profiles using the scaling laws derived for the constant volume propagation in \S 2.2. Experimental parameters: $E = 88$ KPa, $Q_{0} = 10$ ml.min$^{-1}$, $V_{0} = 6$ ml, and $\mu = 10$ Pa.s}
    \label{fig:thick}
\end{figure}
{The time dependence of the radius demonstrates the succession of three regimes of fracture propagation: growth at constant flow rate, growth at constant volume and saturation. The growth of the fracture at constant volume is characterized by its duration and the relative change in fracture radius. The duration and the relative change in radius decrease as the dimensionless toughness of the injection increases \cite{221}. Indeed the larger the value of $\kappa_s$, the closer to the toughness regime the injection is. The fracture is arrested immediately after the injection stops for a cut-off value of $\mathcal{K}_s = 2.5$. For the experiments conducted in this study, the $\mathcal{K}_s$ values have been listed in table \ref{tab:exp} and range between 0.996-2.17, which explains why the growth at constant volume lasts a few minutes for a relative change in radius is about 50$\%$.}

\subsection{Thickness measurements}
To further characterize the three propagation regimes, we measure the fracture aperture with the dye absorption method and plot the profiles of the fractures in figure \ref{fig:thick}. During the injection, both the radius and aperture of the fracture increase. Upon rescaling, the data collapse on a self-similar fracture profile, after an initial transient regime, as shown in figure \ref{fig:thick}(b). After the injection, the radius of the fracture increases as the aperture decreases. The rate of propagation is slower than it was during the injection. Upon rescaling, all data collapse on a second self-similar profile which corresponds to the viscous propagation of a fracture of constant volume (see figure \ref{fig:thick}(c)).

%%%%%%%%%%%%%%%%%%%%%%%%%%%%%%%%%%%%%%%%%%%%%%%%%%%%%%%%%%%%%%%%%%%%%%%%%%%%%%%%%%%
\section{Conclusion}
As a pressurized fluid is injected in an elastic brittle material, a penny-shaped fracture forms and propagates. The complex fracture dynamics depend on the matrix and fluid properties and the injection parameters. During the fluid injection, modeling and experimental studies have demonstrated two asymptotic regimes. The fracture expansion is either controlled by the viscous dissipation in the fluid, in the viscous-dominated regime or the toughness of the material, in the toughness-dominated regime. {Upon shut-in, the continued propagation of the fracture at constant volume has been observed in porous materials and predicted for impermeable matrices. In this study, we experimentally study the propagation regimes of a viscous-dominated fracture in a hydrogel matrix. We demonstrate the existence of three propagation regimes: injection growth, post shut-in propagation, and saturation. For each regime, we show a good agreement with the scaling laws derived for the growth of fractures in the viscous-dominated regime during injection at a constant flow rate and post shut-in. The saturation values of the fracture radius and aperture are reached when the stress intensity factor at the tip of the fracture becomes lower than the material's toughness. The experimental results and corresponding model allow for predicting the relative growth of the fracture after the injection stops, i.e., the final size of the fracture and how long it takes for the fracture to reach this equilibrium geometry once the injection stops.} 

{The datasets supporting this article have been uploaded as part of the supplementary material \cite{34}.}

\section*{Appendix A. Effect of box dimensions}
All experiments are conducted in a 12.5 cm $\times$ 12.5 cm $\times$ 12.5 cm block of gelatin set in an acrylic box of the same dimensions. To test the influence of the box size or edge effects on the fracture dynamics and the saturation radius, we performed experiment 3 (see table \ref{tab:exp} for experimental parameters in a larger volume of gelatin of dimensions 15 cm $\times$ 15 cm $\times$ 15 cm. 
\begin{figure}[h]
    \centering
    \includegraphics[width = 2.5in]{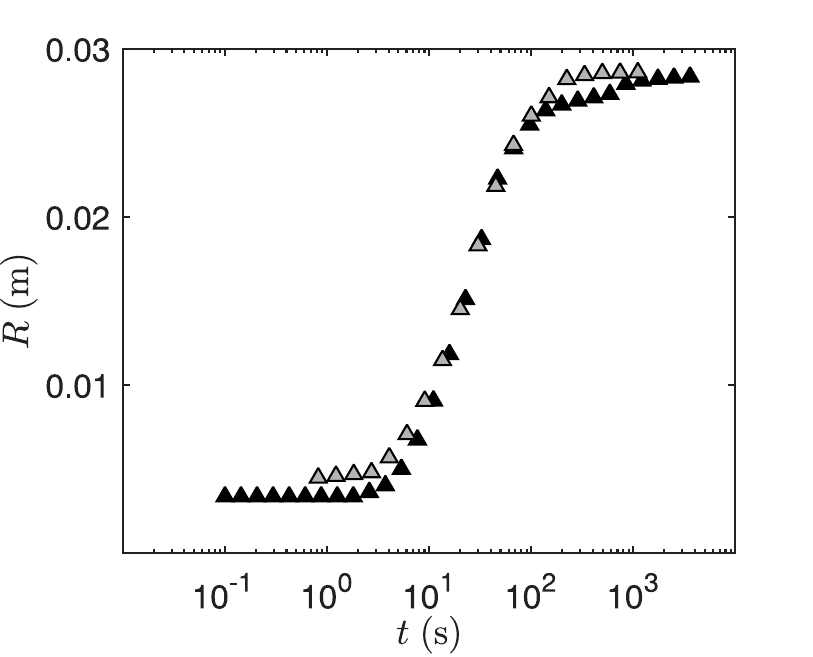}
    \caption{Influence of matrix dimensions on fracture propagation. Radius vs time for experiment 3 (see table \ref{tab:exp} for experimental parameters) conducted in a box of size 12.5 cm $\times$ 12.5 cm $\times$ 12.5 cm (${\color{black} \hspace{0 mm} {\mathlarger{\mathlarger{\blacktriangle}}} }$ ${\hspace{-3.6 mm} \raisebox{-0.5pt}{{$\bigtriangleup$}}}$) and a box of size  15 cm $\times$ 15 cm $\times$ 15 cm (${\color{gray} \hspace{0 mm} {\mathlarger{\mathlarger{\blacktriangle}}} }$ ${\hspace{-3.6 mm} \raisebox{-0.5pt}{{$\bigtriangleup$}}}$). }
    \label{fig:appb}
\end{figure}
The radius of the fracture is recorded over time and plotted in fig.\ref{fig:appb}, for the two block sizes. The results demonstrate that the evolution of the radius does not depend on the size of the block of gelatin. The saturation radius is not set by edge effects. 

\section*{Appendix B. Governing equations}
We review the mathematical derivations that define the radius and aperture of penny-shaped fracture driven by a fluid \cite{21}. 

The mechanical deformation in the elastic matrix associated with the fracture thickness \textit{w} relates to the pressure in the fracture and the fracture radius through the following integral relation, initially derived by Sneddon $\&$ Lowengrub \cite{35}.
\begin{equation}
w(r, t)=\frac{8 R}{\pi E^{\prime}} \int_{r / R}^{1} \frac{\xi}{\sqrt{\xi^{2}-(r / R)^{2}}} \int_{0}^{1} \frac{x p(x \xi R, t)}{\sqrt{1-x^{2}}} \mathrm{~d} x \mathrm{~d} \xi .
\end{equation}
To describe the fluid flow in the fracture, we use the non-linear lubrication equation called the Reynolds equations \cite{36} 
which relates the aperture of the fracture to the pressure and the fracture radius. 
\begin{equation}
    \frac{\partial w(r, t)}{\partial t}=\frac{1}{12 \mu} \frac{1}{r} \frac{\partial}{\partial r}\left(r w^{3}(r, t) \frac{\partial p}{\partial r}\right)
\end{equation}
The stress intensity factor $K_I$ defines the stress concentration at the tip of the fracture. The fracture propagates if the stress intensity factor $K_I$ is equal to $K_IC$, i.e., the material toughness. For a penny-shaped fracture, the stress intensity is equal to \cite{37}: 
\begin{equation}
    K_{I}=\frac{2}{\sqrt{\pi R}} \int_{0}^{R(t)} \frac{p(r, t)}{\sqrt{R^{2}-r^{2}}} r \mathrm{~d} r .
\end{equation}

The boundary conditions are set by the fracture geometry. The integral representation of the fluid mass conservation in the fracture is in equation (\ref{eq:5.5}). The fracture thickness at the tip is 0, equation (\ref{eq:5.6}). There is no flow through the tip of the fracture in the elastic medium, equation (\ref{eq:5.7}).  
\begin{eqnarray}
    Q t=2 \pi \int_{0}^{R(t)} r w(r, t) \mathrm{d} r\label{eq:5.5}\\
    w=0, \quad r=R(t)\label{eq:5.6}\\
    w^{3}(r, t) \frac{\partial p(r, t)}{\partial r}=0, \quad r=R(t)\label{eq:5.7}
\end{eqnarray}

In this study, this set of coupled equations and boundary conditions are used to derive scaling laws for the fracture  radius and aperture.

\end{document}